\documentclass[lettersize,journal]{IEEEtran}

\IEEEoverridecommandlockouts
\usepackage{graphicx, amsmath, amsthm, amssymb, subcaption, url, cite, array, amsthm}
\usepackage[ruled, linesnumbered]{algorithm2e} 
\usepackage{algorithm2e, setspace}
\usepackage[font={small}]{caption}
\usepackage{hyperref}
\usepackage{xcolor}

\makeatletter
\let\@oldmaketitle\@maketitle
\renewcommand{\@maketitle}{
  \@oldmaketitle
  \vspace{-24pt}  
}

\begin{document}
\title{\huge Integrating Brain-Computer Interface and Neuromorphic Computing for Human Digital Twins}

\author{Chen Shang,~\IEEEmembership{Student Member,~IEEE,} Jiadong Yu,~\IEEEmembership{Member,~IEEE,}
Dinh Thai Hoang,~\IEEEmembership{Senior Member,~IEEE,}
\thanks{Chen Shang and Jiadong Yu are with the Internet of Things Thrust, The Hong Kong University
of Science and Technology (Guangzhou), Guangzhou, Guangdong, China (chenshang@hkust-gz.edu.cn, jiadongyu@hkust-gz.edu.cn). Dinh Thai Hoang is with the School of Electrical and Data Engineering, University of Technology Sydney, Australia (hoang.dinh@uts.edu.au).}
}

\maketitle
\begin{abstract}
The integration of immersive communication into a human-centric ecosystem has intensified the demand for sophisticated Human Digital Twins (HDTs) driven by multifaceted human data. However, the effective construction of HDTs faces significant challenges due to the heterogeneity of data collection devices, the high energy demands associated with processing intricate data, and concerns over the privacy of sensitive information. This work introduces a novel biologically-inspired (bio-inspired) HDT framework that leverages Brain-Computer Interface (BCI) sensor technology to capture brain signals as the data source for constructing HDT. By collecting and analyzing these signals, the framework not only minimizes device heterogeneity and enhances data collection efficiency, but also provides richer and more nuanced physiological and psychological data for constructing personalized HDTs. To this end, we further propose a bio-inspired neuromorphic computing learning model based on the Spiking Neural Network (SNN). This model utilizes discrete neural spikes to emulate the way of human brain processes information, thereby enhancing the system's ability to process data effectively while reducing energy consumption. Additionally, we integrate a Federated Learning (FL) strategy within the model to strengthen data privacy. We then conduct a case study to demonstrate the performance of our proposed twofold bio-inspired scheme. Finally, we present several challenges and promising directions for future research of HDTs driven by bio-inspired technologies.

\end{abstract}


\vspace{-10pt}
\section{Introduction}\label{sec-intro}
The paradigm shift in communication networks from 5G to 6G is accelerating the realization of immersive communication. Central to this transformation is the human-centric Metaverse, a shared virtual environment emerging from the seamless fusion of digitally enhanced physical reality, augmented reality (AR), and ubiquitous Internet connectivity. By enabling users to engage in interactive, professional, and social activities within three-dimensional spaces, the Metaverse effectively transcends physical and geographic boundaries. To further bridge the gap between virtual and physical worlds, the concept of human digital twins (HDTs) has emerged. These advanced digital representations not only replicate but also enhance human interactions through sophisticated digital extensions, ensuring that both individual and collective engagements in virtual spaces mirror the depth and nuance of the physical world~\cite{10238695}.


Different from the virtual representation of users in the Metaverse (i.e., avatars), the HDT is an advanced system that extends beyond mirroring physical appearances. This system integrates and analyzes comprehensive data types of user, including behavioral, psychological, physiological, and observational information, enabling the HDT to adapt dynamically not only to user interactions and immersion within the Metaverse but also to provide real-time monitoring of human behavior and states. By leveraging each user's unique and real-time data, HDTs efficiently capture real-time feedback and delivers personalized services, offering tailored solutions that meet the needs of both individual users and Metaverse Service Providers (MSPs). Therefore, this system not only significant enhances the Quality of Experience (QoE) perceived by the users but also assists in optimizing the Quality of Service (QoS) provided by the MSPs, ensuring efficient system performance. These multiple benefits highlight the integral role of HDTs in enhancing the overall functionality and sustainability of the human-centric Metaverse. 

To construct an effective HDT, users must deploy multiple Internet-of-Things (IoT) devices (e.g., smartwatches, smartphones, and body sensors) to continuously collect real-time data on their physiological and behavioral states. This data is then transmitted to an MSP for comprehensive analysis, enabling the creation of highly personalized HDTs. However, this process presents significant challenges in two main stages respectively, i.e., data collection and data processing. Regarding data collection, the IoT devices used in this process are typically produced by different manufacturers, each operating on distinct software platforms and utilizing different standards and protocols. This diversity complicates the efficient integration and processing of data. Varied operational protocols and data formats can lead to inconsistencies and gaps in the collected data, thereby reducing the efficacy of data collection. Additionally, the sensors in these devices are typically designed to detect only specific physiological signals, such as heart rate and blood oxygen levels. However, this may be insufficient in the dynamic real-world environments where human behaviors and states, such as psychological states like emotions and physiological states like fatigue, are complex and variable. Such sensors have limitations in capturing nuanced psychological or emotional states, which are essential for the immersive human-centric ecosystem.

On the other hand, the heterogeneity of IoT devices presents significant challenges for MSPs in efficiently processing data. MSPs must employ sophisticated algorithms capable of handling the diverse types of data generated by these devices. This necessity not only complicates the data integration process, requiring robust solutions to manage different operational protocols and data formats, but also impacts the system's efficiency. Such complexities can diminish the accuracy and responsiveness of HDT, potentially affecting the immersive user experience.
Furthermore, the vast amount of real-time data collected by numerous IoT devices, combined with their rapidly expanding user base, further poses significant pressure on data collection and processing systems while also heightening concerns about energy consumption and data security. 
Specifically, the intensive data handling required not only consumes considerable energy but also imposes strict constraints on the operational efficiency of these systems, thereby limiting the range of practical applications available to users. Although deep learning model are capable of managing these large data volumes by enhancing processing efficiency, their substantial energy requirements and the inherent risks of data breaches pose serious sustainability challenges.

Given these pivotal challenges identified in the development of HDTs, this work aims to answer several key research questions essential for advancing the field:
\begin{enumerate}
    \item[\small{\bf{Q1)}}] How to accurately capture human activities data for constructing a HDT that dynamically adapts to users' psychological and physiological changes, while reducing reliance on heterogeneous devices, thereby improving both the QoS for MSPs and the QoE for users?
    \item[\small{\bf{Q2)}}] Given the need for continuous updating of HDT to adapt to real-time changes of user, what strategies can be developed to reduce the energy consumption involved in these data-intensive operations driven by deep learning model?
    
    \item [\small{\bf{Q3)}}] Given the sensitivity of the individual data involved, how can security measures be enhanced during the continuous updates of HDT to protect against data breaches and efficiently manage vast volumes of sensitive data?
\end{enumerate}
To address these challenges, we revisit current HDT frameworks from the perspective of human-centric Metaverse requirements. We then introduce a novel HDT framework and data processing method driven by biologically inspired (Bio-inspired) technologies. 
Our contributions are summarized as follows: 
\begin{itemize} 
\item  \textcolor{black}{We introduce a novel Bio-inspired HDT (Bio-HDT) framework that utilizes Brain-Computer Interface (BCI) technology to collect electroencephalography (EEG) signals from human brain as the primary data source for constructing HDTs. This framework harnesses EEG signals to capture real-time behavioral, psychological, physiological, and observational information of individuals, thereby providing richer data and a holistic view for constructing immersive and personalized HDTs.} This approach not only enhances the precision and fidelity of the HDTs but also significantly reduces the reliance on heterogeneous IoT devices, simplifying both data collection and processing. Consequently, it ensures high QoS provided by MSPs and QoE perceived by users (\textbf{For Q1}).
\item  We propose a Bio-inspired neuromorphic computing learning model based on the  Spiking Neural Networks (SNNs) for efficiently processing extracted EEG signals. Unlike traditional Artificial Neural Networks (ANNs) that process information continuously and require constant computational activity, SNNs utilize discrete, brain-like spikes to transmit information between neurons. This method more accurately reflects the natural communication processes of brain neurons, thereby enhancing data processing efficiency and reducing the energy consumption specifically involved in the real-time updating of HDTs (\textbf{For Q2}).
\item We introduce an SNN-driven FL model that reduces raw data transmission to address privacy challenge (\textbf{For Q3}). Additionally, we conduct a case study to validate the performance of our proposed Bio-HDT framework and bio-inspired learning model.
\end{itemize}
\textit{To the best of our knowledge, this is the first work to explore the potential of Bio-inspired technologies in tackling the challenges faced by HDTs and the broader human-centric Metaverse.}

\vspace{-8pt}
\section{Human Digital Twin}
In this section, we introduce the human-centric Metaverse and review the current HDT frameworks and their applications. Then, we highlight the key challenges that must be addressed for realizing their full potential within the human-centric Metaverse.
\vspace{-5pt}

\vspace{-5pt}
\subsection{Human-centric Metaverse and Human Digital Twin Framework}
\vspace{-5pt}
As the next generation of the Internet, the human-centric Metaverse extends beyond providing a virtual space for gaming. It is envisioned as a new paradigm that leverages users' behavioral, psychological, physiological, and observational data to bridge the virtual and physical worlds, thereby profoundly transforming human lifestyles. HDTs, similar to the critical role of digital twins (DT) in manufacturing~\cite{rani2024human}, play a vital role in improving the human-centric Metaverse. For instance, they allow simulation and prediction of physiological and psychological states of humans, facilitating personalized medical treatments~\cite{10238695} and health management programs~\cite{9839649}. 


Specifically, the current HDT framework operates cyclically by executing the following steps: (1) Sensors embedded in IoT devices continuously collect user real-time behavioral data; (2) The data are then transmitted to the MSP; (3) The MSP processes and analyzes the data using advanced software and algorithms. After analyzing these data, MSPs are able to generate the user's HDT. Currently, this framework has been utilized primarily in healthcare. For example, HDTs are integrated with wearable devices such as smartwatches, fitness trackers, and health sensors to continuously monitor vital factors like heart rate, blood pressure, and glucose levels. This data can be synchronized with medical records and utilized to generate real-time insights into an individual's health, enabling early diagnosis of potential issues and the optimization of personalized healthcare strategies~\cite{10238695,9839649,asad2023human}.

\vspace{-5pt}
\subsection{Challenges of the Current Human Digital Twins}
While the concept of HDTs offers immense advantages, it is still in its early stages. Multifaceted challenges need to be addressed to fully realize its potential.
\subsubsection{Devices and Data Sources   \textbf{(Q1)}}
One of the significant challenges facing HDTs is the heterogeneity of the IoT devices used to collect human data. These IoT devices such as smartwatches, body sensors, smartphones, and fitness trackers are critical for collecting continuous streams of physiological data. However, these devices are often produced by different manufacturers and have varying technical specifications. They operate on diverse hardware platforms, utilize different communication protocols, and employ distinct software architectures, creating significant barriers to seamless data integration and real-time processing. This heterogeneity complicates the creation of accurate, responsive HDTs, as each device requires specific processing capabilities and tailored integration efforts. Moreover, the need for real-time synchronization of data from multiple sources introduces additional technical complexities, including latency, bandwidth constraints, energy, and data loss, all of which degrade system performance. On the other hand, inconsistent data formats and interoperability issues increase the difficulty for MSPs in developing standardized frameworks for HDT implementation, thereby limiting the scalability and personalization potential of HDTs.

Furthermore, while these existing HDT frameworks can effectively replicate certain physiological functions (e.g., heart rate and blood oxygen), they are still far from achieving the comprehensive and dynamic representation of human bodies and minds that are necessary for fully immersive and personalized experiences in the human-centric Metaverse. The integration of advanced data sources, such as neurological and psychological inputs, remains nascent. This underscores a significant gap in the potential for HDTs to evolve into fully interactive and responsive entities within the human-centric Metaverse.

\subsubsection{Energy Consumption \textbf{(Q2)}} 
To efficiently analyze these collected data, advanced models and algorithms based on deep learning have been deployed. However, they are primarily driven by ANNs, which are resource-intensive and require substantial computational power for both training and inference phases~\cite{wu2018development}.
This intense energy demand presents significant challenges for these IoT devices with limited energy. First, the high power requirement for real-time data processing and analysis can lead to frequent battery depletion, disrupting continuous monitoring and data collection essential for maintaining accurate and dynamic HDTs. Moreover, the strain on device endurance due to persistent high-load operations can degrade device performance over time, increasing maintenance costs and potentially compromising the quality of user interaction within the Metaverse.

\subsubsection{Data Security and Privacy \textbf{(Q3)}}
HDTs are continuously updated with real-time data from IoT devices, introducing inherent risks of data breaches. The complexity of HDTs and the increase in data access points, such as sensors, wearable devices, and edge servers, expand potential cyber threats. Ensuring data security and privacy while maintaining HDT efficiency and responsiveness is a substantial technical challenge.

Based on the analysis above, it is evident that while HDTs have established a solid foundation for their role in certain applications, significant challenges persist. To address these challenges, we propose a novel HDT framework in the next section.

\section{Brain-Computer Interface-Enhanced Human Digital Twin}

In this section, we illustrate BCI technology and its current applications across various domains. We then propose a novel Bio-HDT framework that leverages advanced BCI technology. Finally, we discuss the unique challenges involved in deploying BCI within HDT frameworks.

\subsection{Brain Signal and Brain-Computer Interface}
Unlike previous works~\cite{10238695,9839649,asad2023human}, this work leverages brain signals as a novel data source for constructing HDTs. 
Specifically, brain signals offer a multifaceted view into human nature, crucial for modeling the human activities. They contain critical behavioral, psychological, physiological, and observational information, enabling the decoding of subconscious and conscious responses to environmental stimuli. These insights serve as windows into the internal processes that govern human actions, perceptions, and interactions, presenting their significant potential for constructing HDTs. By harnessing these signals, we can develop a human-centric Metaverse where each user's HDT is authentically driven by their unique brain activities, ensuring a more immersive and personalized virtual experience. 

To this end, BCI technology that connects
the brain signals and physical devices has been considered. As a direct communication pathway between the human brain and externals machines, devices, and system, BCI operates by monitoring brain activities and collecting brain signals through sensors attached to the human scalp (e.g., non-invasive BCI~\cite{cognixion}), processing the data to detect specific patterns related to intended actions, and translating these patterns into actionable commands. This allows users to control devices or perform tasks without physical movement~\cite{neuralink_2024_neuralink}. It has been successfully applied in various domains, such as healthcare in arkinson's~\cite{8119531} and Alzheimer's~\cite{7967704},
demonstrating its broad potential and utility. 
It has been successfully applied across multiple domains, including healthcare for Parkinson's and Alzheimer's disease~\cite{8119531,7967704} and entertainment for thought-driven avatar control and immersive VR interaction~\cite{cognixion,10496440}, demonstrating broad potential and utility.

Unlike standard physical sensors technology that detect straightforward parameters such as temperature, pressure, or light, the human brain processes a rich and complex array of signals related to emotions, cognition, and behaviors. This comprehensive capability enables BCIs to access a spectrum of human states data that conventional sensors typically cannot, e.g., physical, mental, emotional, and behavioral states. Therefore, BCI offers the significant potential to create highly sophisticated and biologically accurate digital representations within the Metaverse, capturing subtleties of the human states that are beyond the reach of physical sensors.

\vspace{-5pt}
\subsection{Biologically-inspired Human Digital Twins}
\begin{figure*}[t]
    \centering
    \includegraphics[width=0.7\textwidth]{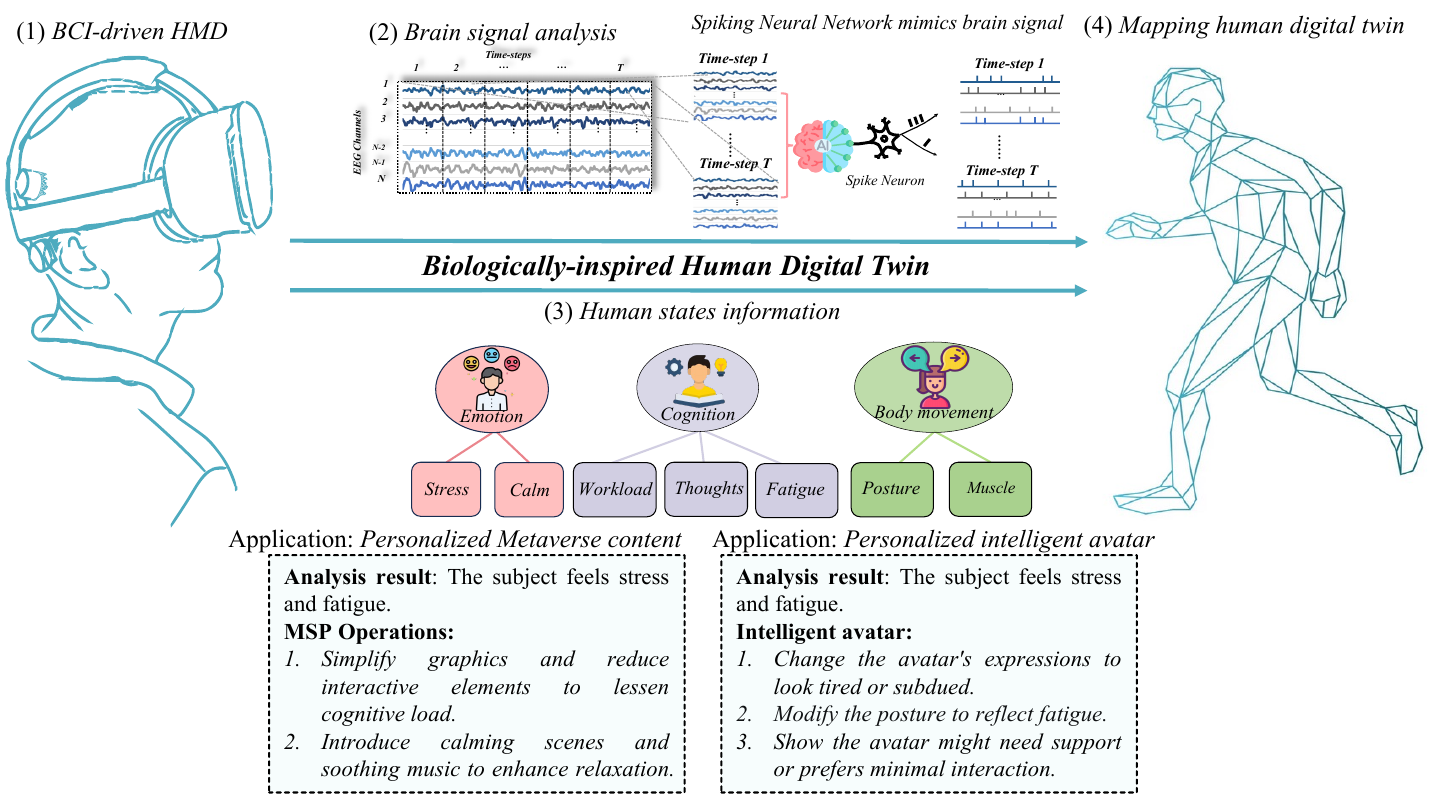}
    \caption{The proposed framework for the biologically-inspired human digital twin (Bio-HDT).}
    \label{fig:Bio-HDT}
    \vspace{-10pt}
\end{figure*}
As analyzed above, brain signals encompass underlying information about human activities. These activities provide a holistic view of an individual's real-time states across multiple dimensions, which is particularly useful in contexts such as healthcare, personalized technology interactions, and immersive Metaverse application. By capturing and analyzing these various human behavior and states information, BCI technology assists in creating more personalized and responsive systems that adapt to individual needs and changes in their condition. 
\textcolor{black}{Therefore, we propose the Bio-HDT framework that utilizes brain signals as the data source for conducting HDT and incorporates BCI technology to efficiently capture these signals. 
In particular, this work primarily focuses on EEG signals due to their ease of non-invasive acquisition, high temporal resolution, and broad applicability in wearable BCI systems~\cite{10496440}.
These characteristics make EEG an ideal modality for real-time monitoring and neural-digital interaction within the proposed Bio-HDT framework.}

As illustrated in Fig.~\ref{fig:Bio-HDT}, the BCI-driven Head-mounted Display (HMD) system within the Bio-HDT framework continuously captures \textcolor{black}{EEG signals from human brain} through embedded non-invasive BCI sensors, which are then analyzed using a neuromorphic computing learning model that mimics brain function through SNNs. After real-time extraction and analysis of \textcolor{black}{these EEG signals}, the Bio-HDT is constructed and synchronized with the corresponding human states in the physical world. Utilizing this real-time Bio-HDT, MSPs can identify and even predict fluctuations in a user's psychology and physical intentions. Consequently, this framework enables MSPs to generate personalized services for the users by \textcolor{black}{monitoring the EEG signals acquired from their brain activity} (\textbf{for Q1}).

Specifically, the Bio-HDT operates as a two-layer bio-inspired scheme. The first layer bio-inspired captures \textcolor{black}{EEG} signals, which reflect the user's cognitive, emotional, and behavioral states, providing an authentic representation of the brain’s biological processes. These \textcolor{black}{EEG} signals are continuously monitored and collected through embedded BCI sensors in the HMD system, ensuring real-time feedback from the user. The second layer bio-inspired processes these \textcolor{black}{EEG} signals using SNNs, a neuromorphic computing scheme that mimics the way neurons communicate in the brain. Different from traditional neural networks, SNNs process data in a manner that mirrors the brain’s natural signaling mechanisms, interpreting neural patterns through the timing and frequency of spikes, which is detailed in Section~\ref{SNN}. This allows MSPs to accurately extract, analyze, and respond to users' brain signals, generating personalized content in real time.

By efficiently combining the BCI technology, the Bio-HDT framework offers several advantages. First, it minimizes the reliance on multiple IoT devices and external sensors traditionally required to monitor a user's physical state. Conventional HDT systems rely on complex sensor networks, resulting in hardware demands, synchronization challenges, and increased energy consumption. By directly leveraging brain signals through the BCI, the Bio-HDT framework gathers critical multifaceted data (e.g., cognition, emotion, and behavior) without relying on an array of external sensors, thereby eliminating inefficiencies, reducing the investment in devices, and lowering computational load and data breach risk. Moreover, these \textcolor{black}{EEG} signals are extracted unconsciously, without requiring explicit physical actions, preserving the user’s immersive experience and providing a more nuanced representation of their state. This facilitates more natural and immersive interactions within the Metaverse. 

Furthermore, the Bio-HDT framework establishes a bi-directional feedback loop between the user and the Metaverse. As users interact with the virtual world, their brain signals influence their HDT's behavior and responses. Conversely, the virtual environment provides stimuli or feedback that affects the user's cognitive and emotional state, creating a continuous interaction between the physical and virtual worlds. As a result, MSPs can provide personalized services based on an individual's unique Bio-HDT. For examples, MSPs can dynamically adapt the appearance, actions, or behavior of an avatar in Metaverse based on the user's real-time physiological or psychological states, i,e., Bio-HDT. Besides, MSPs can also adjust the rendering schemes of VR content such as simplifying graphics and reducing interactive elements within the Metaverse to lessen cognitive load when detecting signs of stress and fatigue in the user.

\begin{figure*}[t]
    \centering
    \includegraphics[width=0.7\textwidth]{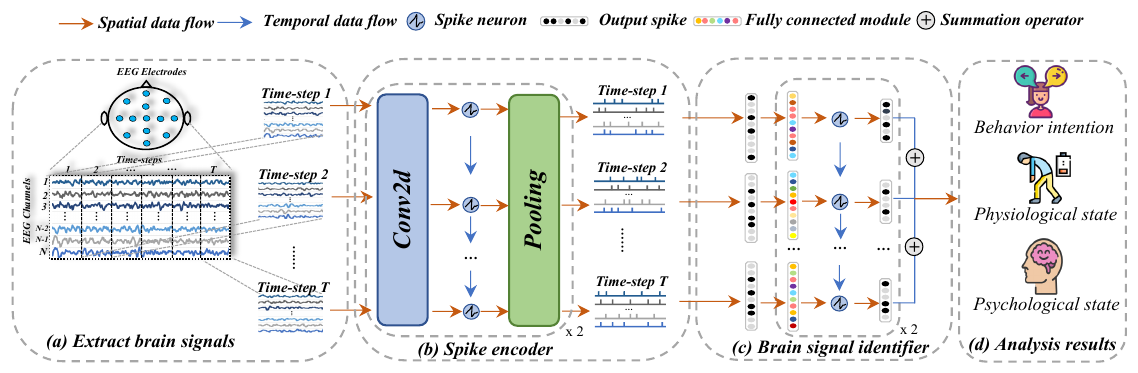}
    \caption{The proposed network architecture based on SNNs. (a) \textcolor{black}{EEG signals extraction} is performed by HMD equipped with EEG electrodes to capture brain activity, (b) The captured brain signal is then input into a spike encoder, which converts the signal into spikes for further processing, (c) and (d) A brain signal identifier processes these firing spikes to accomplish brain signals identification.}
    \label{fig:SNNforHDT}
    \vspace{-5pt}
\end{figure*}

Therefore, the Bio-HDT framework not only addresses the challenge of customizing HDTs for individual users (\textbf{for Q1}) but also minimizes energy consumption (\textbf{for Q2}) and reduces the risk of data breaches \textbf{(for Q3)} by eliminating the need for multiple data collection devices. Additionally, the real-time interaction between brain signals and the decision-making systems enhances the accuracy, responsiveness, and immersiveness of HDTs. This new paradigm brings users closer to a fully realized human-centric Metaverse. It is worth noting that the proposed Bio-HDT system can also serve humans effectively beyond the Metaverse environment. In such cases, the Bio-HDT system can be used to monitor various aspects of human states~\cite{neuralink_2024_neuralink}.

Although the BCI-based Bio-HDT offers rich signals for updating HDTs and advancing the human-centric Metaverse, practical deployment faces three bottlenecks: (i) decoding non-stationary brain activity, (ii) meeting tight energy budgets on always-on edge devices, and (iii) protecting sensitive neural data. These constraints call for efficient, secure, and scalable real-time brain-data pipelines. In the next section, we introduce an SNN-driven FL model that strengthens the Bio-HDT framework by improving dynamic-signal decoding, reducing energy via event-driven computation, and preserving privacy by keeping raw EEG on device.

\section{Spiking Neural Network Enhanced Bio-HDT}\label{SNN}
In this section, we detail the SNN and introduce the SNN-driven FL to enhance the Bio-HDT framework. We then present a case study to support our proposed bio-inspired scheme.
\vspace{-6pt}
\subsection{Spiking Neural Network}
SNN is a bio-inspired neuromorphic computing model that closely mimics the way of human brain processes information. Different from traditional neural networks (i.e., ANNs) that process information continuously and require constant computational activity, SNNs leverage discrete spikes of electrical activity to transmit information between neurons, more accurately reflecting the communication process of biological neurons~\cite{9543525}.
In the context of the Bio-HDT framework and the human-centric Metaverse, SNNs offer significant advantages due to their ability to process information based on the timing and frequency of spikes, closely mirroring the brain's natural information processing mechanisms. These capabilities enables SNNs particularly adept at interpreting rapidly fluctuating brain signals and recognizing dynamic, time-based patterns that traditional models often overlook. As illustrated in the step (2) of Fig.~\ref{fig:Bio-HDT}, extracted brain signals are first fed into a brain-like SNN model. This model learns to process and efficiently encode brain activity into multiple spikes, enhancing the real-time responsiveness and accuracy in user states analysis and personalized service delivery. Importantly, the event-driven nature of SNNs, i.e., activating only during specific spike events, significantly reduces energy consumption and computational overhead, enabling SNNs ideally suited for power-sensitive devices such as HMDs and sensors within the Bio-HDT framework \textbf{(for Q2)}. More technical details of SNN can be found in~\cite{shang2025energy}.

\textcolor{black}{In addition to efficiently processing brain signals, the flexibility of the proposed neuromorphic architecture allows it to be seamlessly extended to other data domains beyond BCI, including physiological, behavioral, and environmental modalities. By converting these diverse inputs into a unified spike-based representation, the SNN model can efficiently learn subtle temporal dynamics across heterogeneous data sources. This generalization enriches the Bio-HDT framework with more comprehensive human-state information and paves the way for multimodal HDT systems in future B5G/6G networks.}

\vspace{-5pt}
\subsection{SNN-Driven Federated Learning}
MSPs traditionally collect users' data to train deep learning models for states analysis. However, transmitting raw brain signals over wireless systems poses significant privacy concerns due to the sensitive and personal nature of this data. To address this challenge, we introduce the SNN-driven FL model. FL is a distributed machine learning approach where models are trained locally on user devices, and only the necessary updates are shared with a central server. This method eliminates the requirement to transmit raw data over the network, thereby enabling user privacy preservation while still enabling the collaborative training of models~\cite{McMahan2016CommunicationEfficient}.

\textcolor{black}{Specifically, SNN-driven FL can be deployed on local devices, i.e., BCI-driven HMDs. Instead of transmitting raw data to MSPs, each device trains its own SNN model using an FL framework. Once the local models are sufficiently trained, only the model updates (i.e., gradients) are transmitted to the MSP. At the MSP, the trained models are aggregated to improve model performance without exposing the original brain data~\cite{McMahan2016CommunicationEfficient}. By retaining raw brain signals on local devices, the system protects users' cognitive, emotional, and behavioral data from potential security threats during transmission. This architecture not only ensures a higher level of data privacy but also addresses critical concerns about the vulnerability of sensitive brain data (\textbf{for Q3}).}

\vspace{-5pt}
\subsection{Case Study}\label{case study}
We demonstrate the efficacy of our framework through comparison simulations in the following. As aforementioned, the Bio-HDT framework is primarily designed to enhance three key factors: personalization, energy efficiency, and privacy preservation. In the human-centric Metaverse, personalization relies heavily on the accurate interpretation of human's behavior intention and states, ensuring that user interactions are not only immersive but also highly responsive. 
For MSPs, the ability to accurately identify and interpret these signals is crucial. This capability is fundamental to effectively analyzing user requirements and creating personalized services that meet individual needs. The precision in processing these signals directly influences the quality, relevance, and timeliness of the content delivered. Consequently, it impacts both the QoS provided by the MSPs and the QoE perceived by the users. Additionally, energy efficiency is crucial for enabling sustainable and scalable operations in real-time environments, ensuring that the system can operate continuously without excessive resource consumption. Therefore, we evaluate our scheme's performance based on the identification accuracy of brain signals and the energy consumption of neural network operations. Note that privacy is not evaluated in this case study because our SNN-driven FL framework effectively addresses privacy concerns by processing data locally. The settings of simulation are as follows:
\textcolor{black}{\subsubsection{BCI data statement} We leverage the public \textit{EEG Motor Movement/Imagery} dataset~\cite{goldberger2000physiobank} to simulate EEG-signal processing. In particular, this dataset includes experimental results where 109 participants perform or imagine specific actions based on the appearance of a target on the screen. \textcolor{black}{After processing, the BCI signals of each subject consist of 255,680 data samples, with each sample including 64 channels, and more details can be found in~\cite{goldberger2000physiobank}\footnotemark\footnotetext{https://physionet.org/content/eegmmidb}.} Specifically, we consider three users, each capable of performing four actions: imagining movement of the left fist, imagining movement of the right fist, imagining movement of both fists, and imagining movement of both feet. The collected data samples are subsequently divided into training and testing datasets at an 80$\%$ to 20$\%$ ratio.}

\subsubsection{SNN architecture} We conduct the learning model based on the SNN framework as shown in Fig.~\ref{fig:SNNforHDT}. This model comprises a spiking encoder network and a classifier network. The spiking encoder is designed to mimic neuronal communication, extracting features from brain signals and converting them into firing spikes at various time steps. The identifier network processes these spikes to carry out identification tasks specifically tailored to brain signals. Specifically, efficient identification of brain signals enables MSPs to understand users' expectations, thus allowing them to provide personalized services.


\subsubsection{FL training and baselines} We utilize the FedAvg algorithm~\cite{McMahan2016CommunicationEfficient} to conduct the FL framework. For the learning hyper-parameters, the learning rate, batch size and local epoch are set as $\left\{ 0.01, 64, 5 \right\} $. \textcolor{black}{Moreover, we conduct performance comparisons between our proposed method, denoted as FL4BCI-SNN, and two baseline methods: (i) FL4BCI-convolutional neural network (CNN)~\cite{LeCun2015Deep}  and (ii) FL4BCI-long short-term
memory (LSTM)~\cite{LeCun2015Deep}. The FL4BCI-CNN baseline is constructed using a traditional ANN-based CNN architecture and adopts an identical structural configuration to FL4BCI-SNN, comprising two convolutional layers, each followed by batch normalization, ReLU activation, and pooling, and one final linear layer for classification. 
The FL4BCI-LSTM baseline consists of two stacked LSTM layers with a single linear output layer, enabling the model to capture temporal dependencies within BCI signals. This consistent architecture design ensures fair comparison between the SNN-based and ANN-based frameworks in terms of depth and parameter scale. 
}
\begin{figure}[t]
    \centering
    \includegraphics[width=0.98\linewidth]{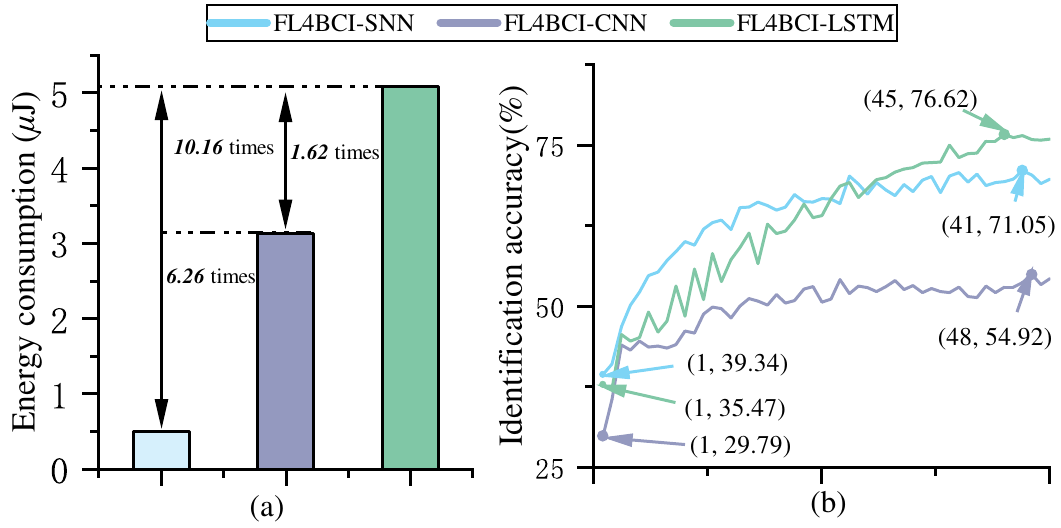}
    \caption{Performance on various methods. (a) Energy consumption of various methods and (b) Identification accuracy vs. training round for various methods.}
    \label{fig:details}
    \vspace{-10pt}
\end{figure}

We compare both energy consumption and identification accuracy between our method and baseline approaches, as shown in Fig.~\ref{fig:details}. 
\textcolor{black}{Following prior work~\cite{shang2025energy}, we consider the model-level compute energy attributable to the neural network per inference step, which can be quantified by decomposing floating-point operations (FLOPs) into primitives.}
In Fig.~\ref{fig:details}(a), we can observe that FL4BCI-SNN exhibits the lowest energy consumption at 0.5$\mu$J, whereas FL4BCI-LSTM records the highest at 5.08$\mu$J, demonstrating an improvement of 10.16 times. Meanwhile, FL4BCI-CNN, which employs an ANN architecture similar to that of FL4BCI-SNN, consumes 3.13$\mu$J. This consumption reflects an energy efficiency gap of more than 6 times compared to FL4BCI-SNN. Besides, as illustrated in Fig.~\ref{fig:details}(b), FL4BCI-SNN shows a progressive improvement in accuracy, reaching 71.05$\%$ over approximately 41 training epochs. In contrast, FL4BCI-CNN achieves a lower final accuracy of 54.92$\%$, demonstrating a substantial gap of 16.13$\%$ below FL4BCI-SNN. FL4BCI-LSTM, outperforms both, achieving the highest accuracy of 76.62$\%$, which is 5.57$\%$ higher than FL4BCI-SNN and 21.70$\%$ higher than FL4BCI-CNN. Despite its superior accuracy, FL4BCI-LSTM also exhibits significantly higher energy consumption, at 5.08$\mu$J, which is approximately 10.16 times greater than that of FL4BCI-SNN at 0.5$\mu$J. This high energy demand poses significant challenges for energy-sensitive wearable devices and sensors. Conversely, compared to FL4BCI-CNN, FL4BCI-SNN not only achieves a 16.13$\%$ improvement in accuracy but also reduces energy consumption by 6.26 times, further demonstrating the efficacy of our scheme.

\vspace{-10pt}
\section{Challenges and Future Research Directions}
In this section, we discuss the Bio-HDT deployment challenges and outlines future research directions.
\vspace{-10pt}
\subsection{Challenges on Processing Brain Signals}

\textcolor{black}{The primary challenge in deploying the Bio-HDT system lies in the accurate and real-time processing of brain signals. Brain signals, often noisy and prone to external interference, exhibit significant variability among users (i.e., neurodiversity). Developing robust signal processing techniques that effectively filter out noise while retaining critical features is essential. Additionally, personalized algorithms may be necessary to account for individual differences, as a generic, one-size-fits-all approach may not be effective across diverse user bases.}

Another critical factor is maintaining real-time synchronization between neural activity and its HDT representation, which is essential for immersion and responsiveness in a human-centric Metaverse. Emerging B5G/6G techniques can reduce end-to-end delay across sensing, inference, and actuation, enabling near-real-time HDT updates. By executing SNN inference and FL aggregation at the edge, the Bio-HDT pipeline can process and respond to EEG streams promptly, thereby deepening interaction fidelity while preserving user experience.

\vspace{-5pt}
\subsection{Challenges in Implementing SNN-driven FL}
One of the primary challenges in deploying SNN-driven FL within the Bio-HDT system lies in the training complexity of SNNs. Unlike traditional ANNs, which process continuous signals, SNNs rely on discrete spikes of electrical activity to transmit information. This discrete, event-driven nature introduces unique difficulties during training. Conventional optimization techniques, such as gradient-based methods (i.e., backpropagation), are less effective for SNNs due to the discontinuous nature of the spike function. Although several approaches, such as surrogate gradient techniques and ANN-to-SNN conversion \cite{9543525}, have been proposed to address these challenges, further exploration is essential to fully unlock the potential of SNNs in real-world applications.

Another significant challenge is the neurodiversity of of brain signals. This variability leads to significant differences in the data across devices, complicating the aggregation and training of a coherent global model in FL. One potential solution is to design cross-learning algorithms that enable efficient transfer learning, allowing the model to adapt to diverse brain signals and improve generalization across users with varying neural characteristics.
\vspace{-5pt}
\section{Conclusion}

This paper proposed the bio-inspired scheme to address the challenges in the construction of HDT. Specifically, we have introduced BCI as the data sources to create the Bio-HDT, which provides richer data and a holistic view for constructing personalized HDTs, while simultaneously reducing the dependency on the diverse data collection devices. We have then proposed a neuromorphic computing learning model that improves the accuracy of brain signals identification and reduces energy consumption by mimicking the brain's natural processing methods, thereby enhancing the system performance. Furthermore, we have proposed the SNN-driven FL to enhance privacy conservation and conducted a case study to validate the efficacy of our proposed scheme. 
Finally, we have discussed some challenges that are worthy of further research. We believe that this work provides valuable insights into the application of bio-inspired techniques in developing personalized immersive services, ultimately advancing the construction of a human-centric next-generation Internet.
\bibliography{bibRef}

@ARTICLE{10238695,
  author={Chen, Jiayuan and Yi, Changyan and Okegbile, Samuel D. and Cai, Jun and Shen, Xuemin},
  journal={IEEE Commun. Surv. Tutor.}, 
  title={Networking Architecture and Key Supporting Technologies for Human Digital Twin in Personalized Healthcare: A Comprehensive Survey}, 
  year={2024},
  volume={26},
  number={1},
  pages={706-746},
  doi={10.1109/COMST.2023.3308717}}

@misc{neuralink_2024_neuralink,
  author = {Neuralink},
  title = {Neuralink},
  url = {https://neuralink.com/.},
  year = {2024},
  organization = {neuralink.com}
}

@ARTICLE{10496440,
  author={Zhu, Howe Yuan and Hieu, Nguyen Quang and Hoang, Dinh Thai and Nguyen, Diep N. and Lin, Chin-Teng},
  journal={IEEE Commun. Surv. Tutor.}, 
  title={A Human-Centric {M}etaverse Enabled by {Brain-Computer Interface}: A Survey}, 
  year={2024},
  volume={26},
  number={3},
  pages={2120-2145},
  doi={10.1109/COMST.2024.3387124}}

@ARTICLE{8119531,
  author={Ju, Ronghui and Hu, Chenhui and zhou, pan and Li, Quanzheng},
  journal={IEEE/ACM Trans. Comput. Biol. Bioinf.}, 
  title={Early Diagnosis of {Alzheimer}'s Disease Based on Resting-State Brain Networks and Deep Learning}, 
  year={2019},
  volume={16},
  number={1},
  pages={244-257},
  doi={10.1109/TCBB.2017.2776910}}

@ARTICLE{7967704,
  author={Mohammed, Ameer and Zamani, Majid and Bayford, Richard and Demosthenous, Andreas},
  journal={IEEE Trans. Neural Syst. Rehabil. Eng.}, 
  title={Toward On-Demand Deep Brain Stimulation Using Online {Parkinson}’s Disease Prediction Driven by Dynamic Detection}, 
  year={2017},
  volume={25},
  number={12},
  pages={2441-2452},
  doi={10.1109/TNSRE.2017.2722986}}

@misc{cognixion,
  title = {Cognixion {One}: The world's first brain computer interface with augmented reality wearable speech generating device},
  url = {https://one.cognixion.com/.},
  organization = {Cognixion ONE}
}

@article{goldberger2000physiobank,
  title={PhysioBank, PhysioToolkit, and PhysioNet: components of a new research resource for complex physiologic signals},
  author={Goldberger, Ary L and Amaral, Luis AN and Glass, Leon and Hausdorff, Jeffrey M and Ivanov, Plamen Ch and Mark, Roger G and Mietus, Joseph E and Moody, George B and Peng, Chung-Kang and Stanley, H Eugene},
  journal={Circulation},
  volume={101},
  number={23},
  pages={e215--e220},
  year={2000},
  publisher={Am Heart Assoc}
}

@ARTICLE{9543525,
  author={Wu, Jibin and Xu, Chenglin and Han, Xiao and Zhou, Daquan and Zhang, Malu and Li, Haizhou and Tan, Kay Chen},
  journal={IEEE Trans. Pattern Anal. Mach. Intell.}, 
  title={Progressive Tandem Learning for Pattern Recognition With Deep Spiking Neural Networks}, 
  year={2022},
  volume={44},
  number={11},
  pages={7824-7840},
  doi={10.1109/TPAMI.2021.3114196}}

@ARTICLE{9839649,
  author={Okegbile, Samuel D. and Cai, Jun and Niyato, Dusit and Yi, Changyan},
  journal={IEEE Netw.}, 
  title={Human Digital Twin for Personalized Healthcare: Vision, Architecture and Future Directions}, 
  year={2023},
  volume={37},
  number={2},
  pages={262-269},
  doi={10.1109/MNET.118.2200071}}

@article{wu2018development,
  title={Development and application of {Artificial Neural Network}},
  author={Wu, Yuchen and Feng, Junwen},
  journal={Wireless Personal Communications},
  volume={102},
  pages={1645--1656},
  year={2018},
  publisher={Springer}
}

@inproceedings{McMahan2016CommunicationEfficient,
  author={H. B. McMahan and E. Moore and D. Ramage and S. Hampson and B. Ag{\"u}era y Arcas},
  title={Communication-efficient learning of deep networks from decentralized data},
  booktitle={Proc. Int. Conf. Artif. Intell. Stat.},
  year={2016},
}

@ARTICLE{shang2025energy,
  author={Shang, Chen and Yu, Jiadong and Hoang, Dinh Thai},
  journal={IEEE Transactions on Wireless Communications}, 
  title={Energy-Efficient and Intelligent {ISAC} in {V2X} Networks with Spiking Neural Networks-Driven {DRL}}, 
  year={2025},
  volume={},
  number={},
  pages={1-1},
  doi={10.1109/TWC.2025.3589438}}

@article{LeCun2015Deep,
  author    = {Yann LeCun and Yoshua Bengio and Geoffrey Hinton},
  title     = {Deep Learning},
  journal   = {Nature},
  volume    = {521},
  number    = {7553},
  pages     = {436--444},
  year      = {2015},
  doi       = {10.1038/nature14539}
}

@article{rani2024human,
  title={A human--machine interaction mechanism: additive manufacturing for Industry 5.0-design and management},
  author={Rani, Sunanda and Jining, Dong and Shoukat, Khadija and Shoukat, Muhammad Usman and Nawaz, Saqib Ali},
  journal={Sustainability},
  volume={16},
  number={10},
  pages={4158},
  year={2024},
  publisher={MDPI}
}

@article{asad2023human,
  title={Human-centric digital twins in industry: A comprehensive review of enabling technologies and implementation strategies},
  author={Asad, Usman and Khan, Madeeha and Khalid, Azfar and Lughmani, Waqas Akbar},
  journal={Sensors},
  volume={23},
  number={8},
  pages={3938},
  year={2023},
  publisher={MDPI}
}

\end{document}